\documentclass[twocolumn,longbib]{aastex7}

\usepackage{CJK}

\hypersetup{pdfauthor={Name}}
\begin{document}
\begin{CJK*}{UTF8}{gbsn}

\DeclareRobustCommand{\okina}{%
  \raisebox{\dimexpr\fontcharht\font`A-\height}{%
    \scalebox{0.8}{`}%
  }%
}

\title{Spectroscopic Characterization of Interstellar Object 3I/ATLAS: Water Ice in the Coma}

\author[0000-0002-5033-9593]{Bin Yang}
\affiliation{Instituto de Estudios Astrof\'isicos, Facultad de Ingenier\'ia y Ciencias, Universidad Diego Portales, Santiago, Chile}
\affiliation{Planetary Science Institute, 1700 E Fort Lowell Rd STE 106, Tucson, AZ 85719, United States}
\email{bin.yang@mail.udp.cl} 

\author[0000-0002-2058-5670]{Karen J. Meech}
\affiliation{Institute for Astronomy, University of Hawai\okina i, 
2680 Woodlawn Drive, Honolulu, HI 96822, USA}
\email{meech@hawaii.edu}  

\author{Michael Connelley}
\affiliation{Institute for Astronomy, University of Hawai\okina i, 
2680 Woodlawn Drive, Honolulu, HI 96822, USA}
\email{msc@ifa.hawaii.edu} 

\author[0000-0003-4936-4959]{Ruining Zhao}
\affiliation{South America Center for Astronomy, National Astronomical Observatories, Chinese Academy of Sciences, Beijing 100101, China}
\affiliation{Instituto de Estudios Astrof\'isicos, Facultad de Ingenier\'ia y Ciencias, Universidad Diego Portales, Santiago, Chile}
\affiliation{School of Astronomy and Space Science, University of Chinese Academy of Sciences, Beijing 100049, China}
\email{rnzhao@nao.cas.cn}

\author[0000-0002-2021-1863]{Jacqueline V. Keane}
\altaffiliation{Research Affiliate University of Hawai\okina i}
\affiliation{U.S. National Science Foundation, 2415 Eisenhower Avenue, Alexandria, VA 22314, USA}
\email{jvkeane@hawaii.edu}

\begin{abstract}
We present optical and near-infrared spectroscopy of the interstellar object 3I/ATLAS, obtained with Gemini-S/GMOS and NASA IRTF/SpeX on 2025 July 5 and 14. The optical spectrum shows a red slope of $\sim$11\% per 1000~\AA{} between 0.5 and 0.8~$\mu$m, resembling typical D-type asteroids and distinct from ultrared trans-Neptunian objects. At longer wavelengths, the near-infrared continuum flattens to $\sim$3\% per 1000~\AA{} between 0.9 and 1.5~$\mu$m, with a broad absorption feature near 2.0~$\mu$m indicative of water ice grains in the coma. Spectral modeling with a mixture of 63\% amorphous carbon and 37\% 1~$\mu$m-sized water ice reproduces both the continuum and the 2.0~$\mu$m band, while the 1.5~$\mu$m water ice band is not detected, likely due to limited signal-to-noise in the IRTF data and dilution by refractory material. The close agreement between the GMOS and SpeX spectra, taken nine days apart, indicates short-term stability in the coma's optical properties. These observations demonstrate that 3I/ATLAS is an active interstellar comet containing abundant water ice, consistent with the theoretical expectation that its home planetary system had a high bulk fraction of water ice by mass.

\end{abstract}


\section{Introduction}\label{sec:intro}

The third interstellar object (ISO) was discovered on 2025 July 1 by the ATLAS Chile telescope \citep{Tonry2018} during its robotic sky survey and was initially designated A11pl3Z. This was a particularly challenging detection because the object lay in Sagittarius, near the galactic plane, a region often avoided by most surveys due to stellar crowding \citep{2025MPEC}. Following the discovery, numerous observatories quickly conducted follow-up observations, and archival searches uncovered pre-discovery images that extended the orbital arc back to 2025 May 22 \citep{Feinstein:2025}. Early measurements indicated an orbital eccentricity greater than 6, and reports of coma or activity prompted its official designation as 3I/ATLAS = C/2025 N1 (ATLAS).

The initial characterization studies of 3I/ATLAS \citep{Seligman2025,delaFuenteMarcos:2025,Bolin:2025}, highlighted both similarities and differences relative to the first two known interstellar objects. Whereas 1I/ʻOumuamua and 2I/Borisov were both estimated to have sub-kilometer radii \citep{Fitzsimmons2024}, early imaging with the Hubble Space Telescope constrains the nucleus diameter to 0.44--5.6 km \citep{Jewitt:2025}. Unlike 1I/ʻOumuamua, which showed no detectable dust coma during its brief observation window, both 2I/Borisov and 3I/ATLAS were discovered with prominent dust comae. Nevertheless, 1I must also have been outgassing, as inferred from the non-gravitational acceleration detected in its trajectory \citep{Micheli2018}. Determining the nucleus rotation has been challenging due to the dense galactic plane background and the presence of a dust coma. However, one study reports a low-amplitude light curve with $\Delta$m $\sim$ 0.08 mag and a period near 17 hours \citep{delaFuenteMarcos:2025}.

James Webb Space Telescope (JWST) Near Infrared Spectrograph (NIRSpec) spectroscopy at $r_\mathrm{H}=3.32,\mathrm{au}$ shows a $\mathrm{CO_2}$-dominated coma with $\mathrm{CO_2}/\mathrm{H_2O}=8.0\pm1.0$, and detections of $\mathrm{H_2O}$, CO, OCS, water ice, and dust \citep{Cordiner:2025}. Swift/UVOT detected OH near $3085,\text{\AA}$, giving $Q(\mathrm{H_2O})=(1.36\pm0.35)\times10^{27},\mathrm{s}^{-1}\approx40\ \mathrm{kg},\mathrm{s}^{-1}$ at $r_\mathrm{H}=3.51,\mathrm{au}$, consistent with activity from icy grains \citep{Xing:2025}. Optical spectra obtained two days after discovery showed a red, nearly featureless reflectance, with no strong gas bands at that time, indicate red dust, with reported spectral slopes of $S'$ = 17--19\% per 1000~\AA~between 0.4--0.8 $\mu$m\citep{Opitom2025,Seligman2025} later VLT/UVES and X-shooter data revealed emergent CN emission and numerous Ni I lines at $r_\mathrm{H}\simeq2.85\text{--}3.2,\mathrm{au}$ \citep{Rahatgaonkar:2025}. An assessment of 3I/ATLAS within the \=Otautahi–Oxford model infers, from its kinematics, an age exceeding 7.6 Gyr and a high H$_2$O mass fraction \citep{Hopkins:2025}.

In this paper, we present spectroscopic observations of 3I/ATLAS aimed at characterizing its dust coma.

\section{Observations and Data Reduction}\label{sec:obs}

\subsection{Gemini}\label{sec:gemini}

The Gemini Multi-Object Spectrograph (GMOS) on the Gemini South telescope \citep{Crampton:2000} was used to obtain low resolution spectra of 3I/ATLAS on 2025 July 4 and 5 UT. Triggering a rapid target of opportunity program [GS-2025A-Q-113], we obtained 6 exposures using the R150 grating and a 1.0$''$ slit, with central wavelength at 0.6 $\mu$m. We used a spatial binning of 2$\times$2 and spectral binning 4$\times$4. We also observed two G2V solar analog stars (HD 174466 \& HD 167221) at airmasses similar to 3I, serving as both telluric correction standards and as solar analog references. GMOS data were calibrated with bias frames, internal spectroscopic flats, and CuAr arc lamp exposures taken close in time to the science frames for the wavelength solution. The data were reduced using the DRAGONS (Data Reduction for Astronomy from Gemini Observatory North and South) pipeline \citep{Labrie:2019}, the official data reduction package provided by the Gemini Observatory.

\begin{deluxetable*}{cccccccc}
\tablenum{1}
\tablecaption{Log of observations.\label{tab:obslog}}
\tablewidth{0pt}
\tablehead{
    \colhead{Observation Date} & \colhead{Tel./Inst.} & \colhead{Object\tablenotemark{$\dagger$}} & \colhead{Exposure} & \colhead{Airmass} & \colhead{$r_{\rm h}$\tablenotemark{a}} & \colhead{$\Delta$\tablenotemark{b}} & \colhead{($^{\circ}$)}\\
    \colhead{(UT)} & \colhead{} & \colhead{} & \colhead{(s)} & \colhead{} & \colhead{(au)} & \colhead{(au)} & \colhead{Phase}
}
\decimals
 \startdata
    2025-07-04  & Gemini/GMOS   & 3I        & 6$\times$200 & 1.26--1.35 & 4.404    & 3.408    & 2.96\\
    $\cdots$ & $\cdots$      & HD 174466  & 4$\times$3   & 1.10--1.11 & $\cdots$ & $\cdots$ & $\cdots$\\
    \hline
    2025-07-05  & Gemini/GMOS & 3I         & 6$\times$200  & 1.37--1.50 & 4.371    & 3.379    & 2.65\\
    $\cdots$ & $\cdots$ & HD 167221  & 4$\times$3    & 1.23--1.25 & $\cdots$ & $\cdots$ & $\cdots$\\
    \hline
      $\cdots$  & $\cdots$    & HD 156734  & 8$\times$15  & 1.46--1.48 & $\cdots$ & $\cdots$ & $\cdots$\\
    2025-07-14  & IRTF/SpeX  & 3I       & 28$\times$120 & 1.27--1.36 & 4.072    & 3.143    & 6.60\\
    $\cdots$ & $\cdots$  & HD 167221   & 16$\times$12  & 1.25--1.26 & $\cdots$ & $\cdots$ & $\cdots$\\
    \enddata
\tablenotetext{\dagger}{All solar analogs are G2V stars, as verified in SIMBAD. }
\tablenotetext{a}{Heliocentric distance}
\tablenotetext{b}{Geocentric distance}
\tablenotetext{c}{Phase angle}
\end{deluxetable*}

\subsection{IRTF}\label{sec:IRTF}
Observations were conducted on the evening of 2025 July 14 (UT) using Director's Discretionary Time (DDT) with the SpeX spectrograph \citep{Rayner2003} on the NASA Infrared Telescope Facility (IRTF). We conducted observations using the SpeX instrument in combination with the MIT Optical Rapid Imaging System (MORIS) camera \citep{Gulbis:2011}. We employed the 0.7\,$\mu$m dichroic in SpeX, which transmits near-infrared light to the spectrograph while reflecting visible light ($\lambda \lesssim 0.7\,\mu$m) to MORIS. Instead of using the facility guiding system (GuideDog), we guided directly on the target using MORIS. This setup enabled stable guiding of the faint interstellar object in a crowded field, while simultaneously acquiring near-infrared spectra with SpeX. NIR data were acquired in prism mode covering the 0.8--2.5 $\mu$m wavelength range, using a 0.8$''$ $\times$ 15$''$ slit, which provided a spectral resolving power of R$\sim$100 \citep{Rayner2003}. Two nearby G2V solar analog star observed along with 3I, served both as the telluric correction standard and as a solar analog reference. Both the science target and the solar analogs were observed at the parallactic angle. Following each IRTF/SpeX science sequence, we obtained internal flat fields and arc lamp exposures for wavelength calibration. The spectral data were reduced using the SpeXtool data reduction pipeline \citep{Cushing2004}.
A journal of the observational details and target geometry is provided in Table 1.

\section{Observational Analysis}\label{sec:observations}
\begin{figure}[h!]
\centering
\includegraphics[width=1.\linewidth]{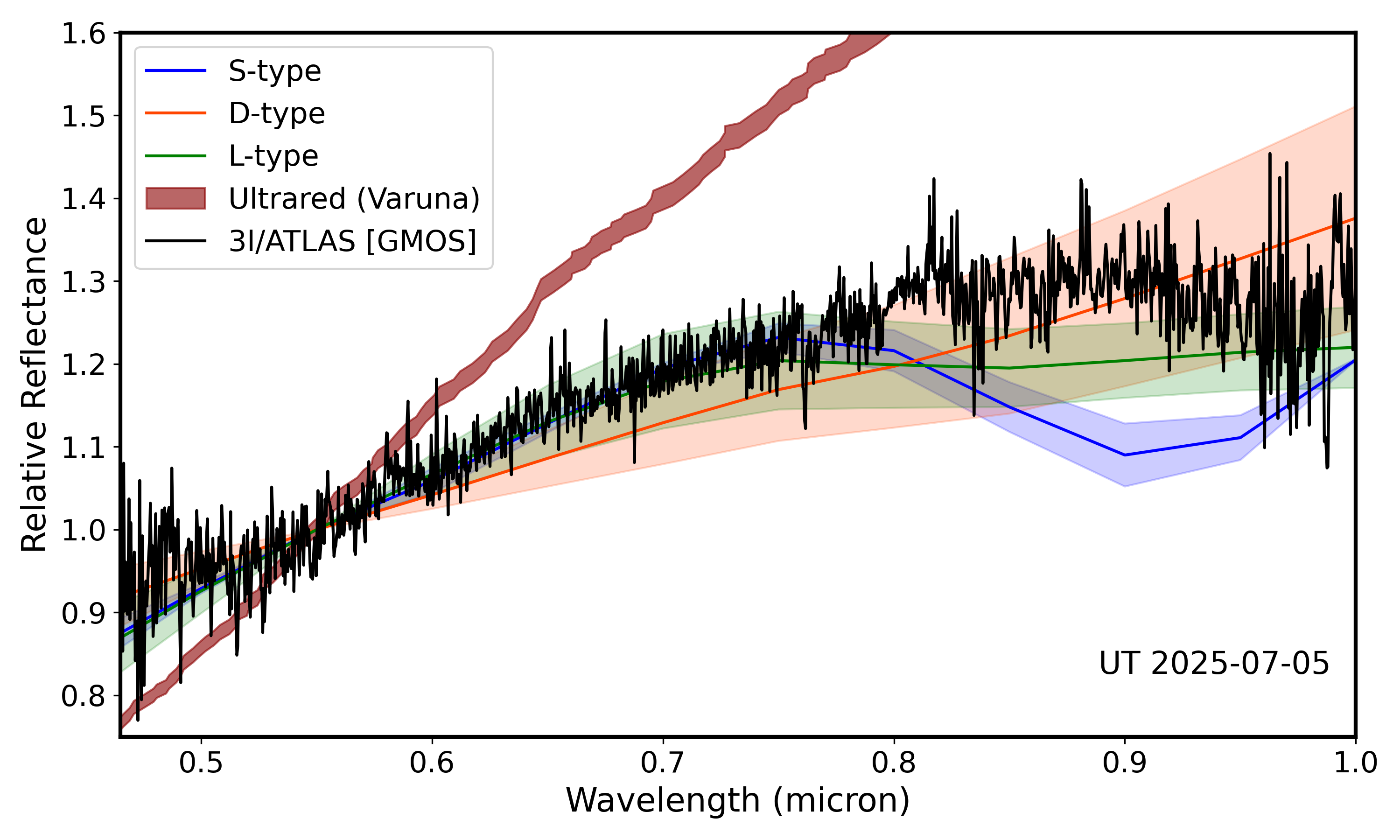}
\hspace{-0.1cm}
\caption{Visible-wavelength reflectance spectrum of 3I/ATLAS obtained with GMOS instrument on Gemini-S, normalized at 0.55 $\mu$m. The colored lines and shaded regions represent the mean and 1$\sigma$ ranges for three major asteroid spectral types: S-type (blue), D-type (red), and L-type (green), taken from \cite{DeMeo2009}. The solid colored lines indicate the average reflectance for each class, while the surrounding shaded regions show the typical variability (1$\sigma$) within each class. The dark red spectrum is (20000) Varuna, a large trans-Neptunian object in the Kuiper belt, representative of ultrared objects in the outer Solar System, taken from \citep{Fornasier:2009}. The comparison shows that 3I/ATLAS has a featureless, red spectral slope, consistent with D-type asteroids; its coma shows no evidence for ultrared matter. }
\label{fig:gmos_spec}
\end{figure}

The GMOS spectrum of 3I/ATLAS, normalized at 0.55~$\mu$m, is shown in Figure \ref{fig:gmos_spec}. It is largely featureless and exhibits a moderately red spectral slope. The comet was observed on two nights, 2025 July 4 and 5, each using a different solar analog for calibration. The July 4 data were heavily contaminated by background stars, leaving only one out of six exposures usable. Although the July 4 data have a much lower signal-to-noise ratio (SNR) than those from July 5, the spectral profile and slope are in good agreement across both nights. The measured spectral slopes over the 0.5--0.8\,$\mu$m range are 14.1~$\pm$~1.3\,\% per 1000\,\AA\ (July~4) and 11.4~$\pm$~0.2\,\% per 1000\,\AA\ (July~5). Due to the contamination and limited usable signal, we exclude the July~4 data from further analysis and focus on the higher-quality July~5 dataset.

To place 3I/ATLAS in the context of known Solar System populations, we compared its reflectance spectrum to representative asteroid taxonomic classes and to ultrared Kuiper belt objects \citep{Jewitt:2002}, shown in Figure \ref{fig:gmos_spec}. We note that because the spectrum of 3I is coma dominated, the spectral slope need not reflect composition alone; it is also shaped by the grain size distribution in the coma. We found that the 3I spectrum is broadly similar to that of D-type asteroids with a moderately red spectral slope, but it does not exhibit the ultrared optical-NIR colors seen in some trans-Neptunian objects such as Varuna, further distinguishing its coma from those of the most compositionally extreme outer solar system bodies. However, the red slope in the GMOS data is not continuous; instead, the spectrum shows a slight downturn, resulting in a flattened slope beyond 0.85 $\mu$m. This downturn may indicate the presence of silicate minerals, which exhibit diagnostic absorption features near 1$\mu$m, commonly observed in S-type asteroids, shown as the green line in Figure \ref{fig:gmos_spec}. 

\begin{figure}[h!]
\centering
\includegraphics[width=1.\linewidth]{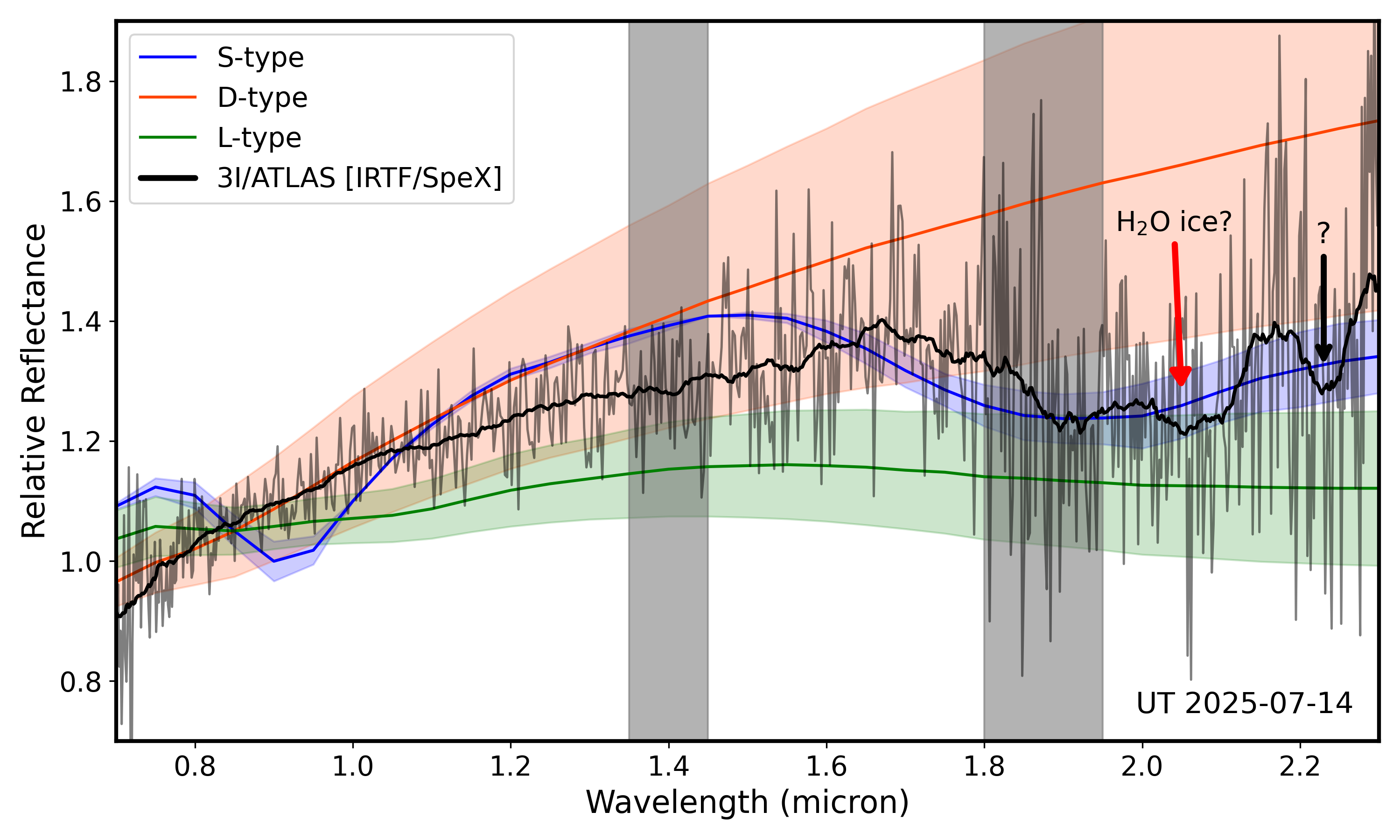}
\hspace{-0.1cm}
\caption{The black line shows the IRTF/SpeX spectrum of 3I/ATLAS, normalized at 0.85 $\mu$m. Colored lines and shaded regions represent the mean and $\pm\sigma$ ranges for S-type (blue), D-type (red), and L-type (green) asteroids, as in Figure \ref{fig:gmos_spec}. A 40-pixel running mean is overplotted on the original IRTF spectrum to illustrate the overall spectral trend. A red arrow marks the broad feature centered near 2.05~$\mu$m, which may be due to water ice. A black arrow indicates a narrow absorption band centered near 2.23~$\mu$m of uncertain origin, possibly related to C-€"H vibrations in organics or hydrocarbons.  Gray regions (1.35-1.45 $\mu$m and 1.80-1.95 $\mu$m) indicate regions of strong telluric absorption from water vapor. }
\label{fig:irtf_spec}
\end{figure}

We requested DDT time on the IRTF to follow up on a possible absorption feature beyond 0.9 $\mu$m, suggested by the shape of the GMOS optical spectrum. Similar to the previous GMOS observations, two nearby G2V stars were observed as solar analogs. The resulting reflectance spectra show comparable profiles. However, the airmass difference between 3I and HD~156734 was larger, leading to a poorer telluric correction. We therefore used HD~167221 to compute the reflectance and for subsequent analysis. The resulting near-infrared spectrum of 3I/ATLAS, obtained with the SpeX instrument, is shown in Figure \ref{fig:irtf_spec}. To compare the GMOS data with the IRTF spectrum, we re-normalized the GMOS to 0.85~$\mu$m and normalized the IRTF spectrum at the same wavelength. In the overlapping region across the 0.7--0.9\,$\mu$m range, the measured GMOS slope is 5.2\,$\pm$\,0.2\,\% per 1000\,\AA{} and the IRTF slope is 6.7\,$\pm$\,1.4\,\% per 1000\,\AA{}. Beyond 0.9~$\mu$m, the GMOS spectrum shows a negative spectral slope of $-3.7 \pm 1.0$ \% per 1000,\AA, whereas the IRTF spectrum maintains the positive slope observed in the optical. To investigate this discrepancy, we compared our data with X-shooter/VLT observations \citep{Alvarez2025} obtained just one day prior to the GMOS dataset. The X-shooter spectrum confirms the continuation of a positive slope from 0.9 to 1.2~$\mu$m, with no evidence of silicate-related absorption features near 1.0~$\mu$m. The apparent negative slope arises at the red edge of the GMOS spectrum, where the instrument sensitivity is lowest; consequently, this portion of the data is unreliable and cannot be interpreted as intrinsic to the object.

To better illustrate the continuum trend, we overplot a 40-pixel smoothed spectrum in Figure~\ref{fig:irtf_spec}. The spectrum of 3I exhibits a broad but shallow absorption feature centered at 2.05~$\mu$m (marked by a red arrow) with a depth of $11.5 \pm 1.7$\%. Such a feature could in principle be attributed to silicates, which commonly produce absorptions near 2.0~$\mu$m in S-type asteroids (shown in blue). However, the absence of the characteristic 1.0~$\mu$m silicate band and the clear mismatch with the S-type spectral profile make a pyroxene origin unlikely. An alternative explanation is water ice, the dominant volatile in most Solar System comets; we examine this possibility in the following section. In addition, a narrow absorption feature is observed between 2.2 and 2.3~$\mu$m (marked by a black arrow), though its origin cannot be unambiguously identified. One possibility is that it may arise from C–H vibrations in organic or carbonaceous materials. Additional observations will be necessary to verify the presence of this feature and better constrain its source.


\subsection{Water Ice Detection and Spectral Modeling}

We generated a synthetic spectral model with amorphous carbon and water ice, computed using Mie theory and following the approach described in \cite{zhao:2025}. 

The real part ($n$) of the refractive index of water ice was taken from \cite{Warren:1984}, and the imaginary part ($k$) from \cite{Grundy:1998}. This approach of combining two different sets of optical constants has been adopted in earlier studies \citep[e.g.,][]{Seelos:2008,Grundy:2009}. Based on the heliocentric distance of 3I/ATLAS at the time of observation, and assuming an albedo of 0.5 for pure water ice and an emissivity of 0.9, the equilibrium temperature of ice grains at 4 au is about 120 K. We verified the robustness of our results by testing alternative optical constant datasets, and found the differences in the modeled spectra to be negligible. We did not choose the widely applied dataset of \cite{Mastrapa2009}, as those optical constants begin at 1.1~$\mu$m and therefore omit the shorter wavelength range still valuable for our analysis. The optical constants of amorphous carbon, used as a representative refractory material, were taken from \citep{Rouleau:1991}.

We assumed that each of the two compositions follows a modified power-law differential size distribution over the range of $a_{\rm min}<a<a_{\rm max}$, i.e., 
\begin{equation}
    \frac{dn(a)}{da}\propto\left(1-\frac{a_{\rm min}}{a}\right)^{\gamma}\left(\frac{a_{\rm min}}{a}\right)^{\alpha}~~,
\end{equation}
where $\alpha$ is the power-law index, and $\gamma$ is related to the peak size $a_{\rm p}$ by $a_{\rm p}=(\gamma/\alpha+1)a_{\rm min}$.

\begin{figure*}
\begin{center}
\includegraphics[width=0.8\linewidth]{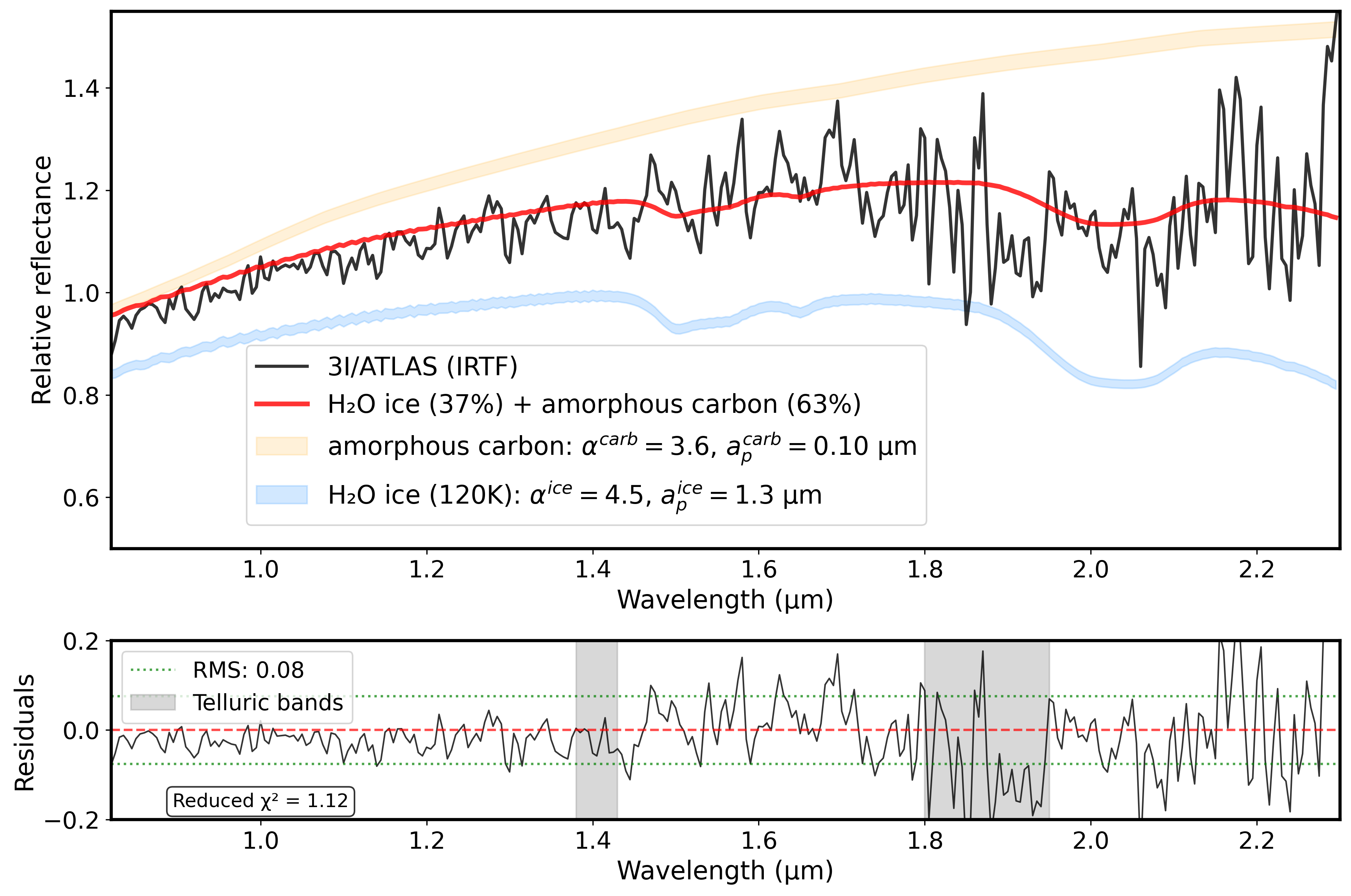}
\end{center}
\caption{Near-infrared spectrum of 3I compared with best-fit spectra model. The black line shows the smoothed reflectance spectrum of 3I/ATLAS, obtained with IRTF/SpeX instrument and normalized at 1.0 $\mu$m. The thick red line represents the best-fit volume mixing model, consisting of 37\% water ice (H$_2$O) and 63\% amorphous carbon. The faint orange and blue bands show the spectral profiles of the two individual component used in the model with different size distribution and peak particle size.}
\label{fig:model}
\end{figure*}

Our best-fit model, shown as the solid red line in Figure \ref{fig:model}, successfully reproduces the 3I/ATLAS spectrum well, capturing both the continuum slope and the broad absorption feature near 2.0 $\mu$m. It consists of a volume mixture of 63\% amorphous carbon and 37\% water ice at a temperature of 120\,K. The carbon grain size distribution follows a power-law with an index of 3.6, peaking at 0.10\,$\mu$m, which is much smaller than the peak size of the water ice grains at 1.3\,$\mu$m. The water ice grain size distribution is characterized by a steeper power-law index of 4.6, indicating a relatively larger population of small grains. We note that, given the limited signal-to-noise ratio (SNR) of the IRTF spectrum and the broad parameter space, our best-fit solution is not unique; other combinations of water-ice fraction and grain-size distribution can yield acceptable, though not optimal, fits to the data. But in any case, the presence of water ice is required to reproduce the absorption feature near 2.0 $\mu$m. 

\begin{figure}
\includegraphics[width=1.\linewidth]{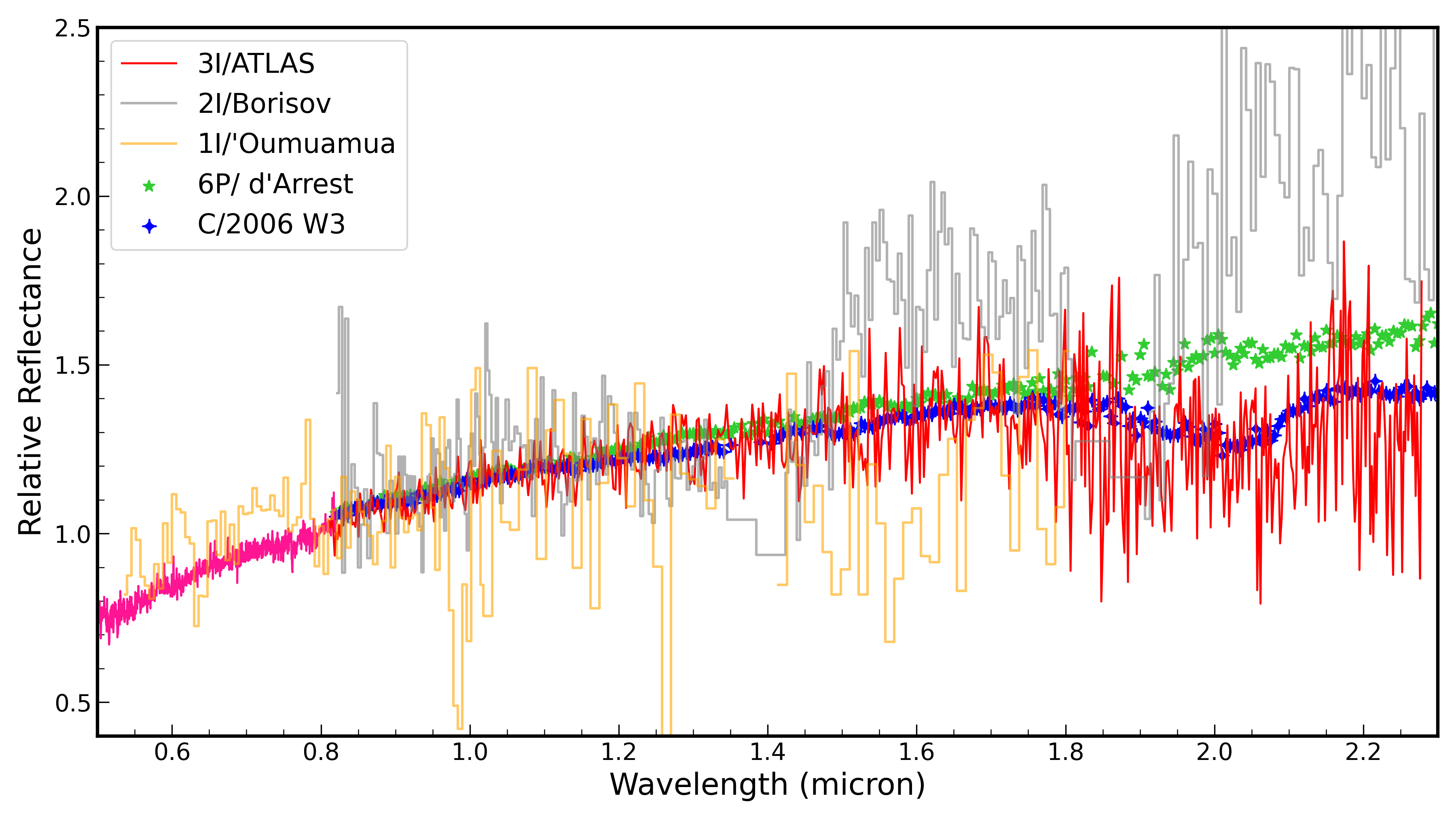}
\caption{Reflectance spectrum of 3I/ATLAS, obtained with Gemini-S GMOS (pink) and NASA IRTF SpeX (red), shown alongside spectra of other known interstellar objects (ISOs) and solar system bodies. Among the ISOs, 3I/ATLAS exhibits the highest SNR and an intermediate spectral slope. In the NIR, its spectrum closely resembles that of comet C/2006 W3 \citep{Yang:2009b}, a bright Oort Cloud comet, with both displaying a moderately red slope and a broad water-ice absorption band at 2.0~$\mu$m.}
\label{fig:3I_colors}
\end{figure}

The diagnostic 1.5 $\mu$m absorption band typically associated with water ice is absent from the 3I spectrum. Such suppression of the 1.5 $\mu$m band has also been reported in the spectra of several active comets known to contain water ice \citep[][also see Figure~\ref{fig:3I_colors}]{Yang:2009a,Yang:2014}. During the {\sl Rosetta mission}, the 2.0 and 3.0 $\mu$m bands were clearly present, while the 1.5 $\mu$m band was weak \citep{Barucci2016}. A similar effect is seen in Centaurs, where water ice is sometimes detected at 2.0 and 3.0 $\mu$m, but the 1.5 $\mu$m band remains weak due to mixing with dark surface materials \citep{Guilbert2009}. The 1.5 $\mu$m water-ice absorption band is often difficult to detect because it is much shallower than the stronger 2.0 and 3.0 $\mu$m bands. At low spectral resolution, the 1.5 $\mu$m feature can be washed out, particularly when the SNR is low or the spectrum is dominated by coma emission. As shown by \cite{Yang:2014}, particle size strongly influences the strength of water-ice absorption features. In our spectral model, relatively small water-ice grains produce only a weak 1.5~$\mu$m band, which could easily be obscured by the noise in the IRTF data. 


\subsection{Comparison to Other Known Objects}

We compared the spectrum of 3I/ATLAS with those of other solar system bodies and known ISOs. All three ISOs exhibit moderately red spectral slopes in the optical and the NIR. Among them, 3I has the highest SNR to date, allowing for tighter constraints on its composition and dust properties. While 2I/Borisov shows a much redder slope in the H and K bands, 3I is notably less red in this region, likely due to the presence of water ice, as indicated by the spectral flattening beyond 1.5 $\mu$m.

The spectrum of 3I closely resembles that of the Oort Cloud comet C/2006 W3 \citep{Yang:2009b}, which likewise lacks the 1.5~$\mu$m absorption feature but shows a strong 2.0~$\mu$m water-ice band. In addition, the optical--NIR spectral slope of 3I is very similar to that of the Jupiter-family comet 6P/d'Arrest \citep{Yang:2009b}. Despite their distinct source regions, 3I and these active Solar System comets exhibit striking spectral similarities. Because the scattering cross section of the dust coma dominates over that of the small nucleus, the observed resemblance likely reflects comparable particle-size distributions or similar dust compositions in their comae.

\section{Discussion}

As the most abundant solid in protoplanetary disks, water ice plays a key role in planet formation \citep{Bitsch:2016,Muller:2021}. In the Solar System, it likely contributed to the cores of giant planets \citep{Ciesla2014} and is a major component of comets \citep{AHearn2017, Mumma:2011}. The detection of water ice in the coma of 3I/ATLAS suggests that it formed beyond the snowline of its parent system, and that water ice could be its major component, similar to comets in the solar system. The derived abundance of 37\% water ice by volume in the coma is consistent with the theoretical study predicting a water-ice mass fraction of 20--40\% for 3I-like objects \citep{Hopkins:2025}. 

The physical structure of water ice records its formation conditions. Below $\sim$100 K, water forms thermodynamically unstable amorphous ice, which gradually converts to crystalline ice above $\sim$ 137 K \citep{Kouchi1994}. Due to limited SNR in the IRTF data, we cannot constrain the ice phase through spectral modeling. IRTF spectra taken 10 days earlier set a $<$7\% upper limit on coma ice \citep{Kareta:2025}, while our observations suggest that water ice accounts for approximately 37\% of the coma by volume. In addition, recent IRTF observations show that the water ice signature appears to be stronger as the comet's activity increases \citep{Lisse:2025}. The increased activity not only enhanced the comet's brightness, improving the SNR of the spectrum, but may also have promoted the ejection of additional icy grains. The JWST/NIRSpec observations clearly show strong water-ice absorption features at 3.0 and 4.5~$\mu$m \citep{Cordiner:2025}. Detailed spectral modeling of these data can provide valuable information on the size, temperature, and impurities of the water-ice grains, and place strong constraints on the formation conditions of this ISO. 

Beyond the absorption feature near 2.0 $\mu$m, the spectral slope of the continuum provides key insights into the dust composition of the coma. Several studies have reported optical spectral slopes for 3I/ATLAS using various instruments and observing conditions, with values ranging from 16\% to 27\% per 1000 \AA\ \citep{Opitom2025, Seligman2025, Puzia:2025, Bolin:2025}. Similar large ranges of measured spectral slope is also reported among previous studies of 1I and 2I. 

Our measurement of $S'$ = 11\% per 1000 \AA, derived from Gemini South/GMOS data taken on July 5, is lower than this range. The observed variation in spectral slope could be due to using different solar analog stars. The near-infrared slope measured with IRTF/SpeX over the 0.8--2.3~$\mu$m range is $1.4 \pm 0.1$\% per 1000~\AA, significantly flatter than the optical slope, and consistent with the presence of water ice in the coma.

Despite being obtained 9 days apart with different instruments, the GMOS and SpeX datasets are statistically consistent in the overlapping 0.7--0.9~$\mu$m region, indicating temporal stability in the continuum shape and confirming the reliability of the optical-NIR transition. This comparison underscores the importance of coordinated, multi-wavelength observations in characterizing transient interstellar objects. Single-instrument datasets may be affected by instrumental or environmental biases, but combined optical and infrared spectra enable a more robust and comprehensive understanding of composition and activity.

\section{Summary}

We conducted optical and near-infrared spectroscopic observations of the interstellar object 3I/ATLAS using Gemini South/GMOS on UT 2025 July 5 and IRTF/SpeX on 2025 July 14. Our main findings are as follows:

\begin{enumerate}
 
\item The optical spectrum of 3I/ATLAS, obtained with GMOS, is featureless and exhibits a moderately red slope of $11.4 \pm 0.2$\% per 1000\,\AA{} between 0.5 and 0.8~$\mu$m. This spectral behavior is comparable to D-type asteroids and active comets in the Solar System, but markedly different from the ultrared objects in the trans-Neptunian region. The measurement is consistent with slopes reported in earlier studies and supports the interpretation that 3I's dust is similar in reflectance properties to Solar System cometary material rather than the ultrared TNO population.

\item The near-infrared spectrum from IRTF/SpeX displays a neutral to slightly bluish slope between 1.7 and 2.3~$\mu$m, together with a broad $\sim$11\% absorption band centered near 2.0~$\mu$m. These features are well reproduced by a spectral model consisting of amorphous carbon mixed with $\sim$1.3~$\mu$m-sized water ice grains, with water ice contributing about 37\% of the coma volume. The flatter continuum relative to the optical, combined with the detection of the 2.0~$\mu$m absorption, provides strong evidence that abundant water ice is present in the coma.

\item The GMOS and SpeX datasets, taken with different instruments and separated by nine days, display consistent spectral slopes in the overlapping 0.7--0.9~$\mu$m region. This agreement indicates that no significant short-term spectral evolution occurred during that period, the overall scattering properties of the coma remained stable over timescales of at least one week.
\end{enumerate}

Our detection of H$_2$O ice in 3I/ATLAS at $\sim$4~au with IRTF, together with the strong CO$_2$ outgassing revealed by JWST and SPHEREx, supports the hypothesis that this object formed in a cold, volatile-rich region of its parent protoplanetary disk. These observations also demonstrate that such volatile compositions can be preserved through ejection and maintained during the subsequent interstellar journey.

\begin{acknowledgments}

Based on observations obtained at the international Gemini Observatory, a program of NSF NOIRLab, which is managed by the Association of Universities for Research in Astronomy (AURA) under a cooperative agreement with the U.S. National Science Foundation on behalf of the Gemini Observatory partnership: the U.S. National Science Foundation (United States), National Research Council (Canada), Agencia Nacional de Investigaci\'{o}n y Desarrollo (Chile), Ministerio de Ciencia, Tecnolog\'{i}a e Innovaci\'{o}n (Argentina), Minist\'{e}rio da Ci\^{e}ncia, Tecnologia, Inova\c{c}\~{o}es e Comunica\c{c}\~{o}es (Brazil), and Korea Astronomy and Space Science Institute (Republic of Korea).
Visiting Astronomer at the Infrared Telescope Facility, which is operated by the University of Hawaii under contract 80HQTR24DA010 with the National Aeronautics and Space Administration. B.Y. is supported by the China-Chile joint research funding CCJRF2209. K.J.M., acknowledges support from the Simons Foundation through SFI-PD-Pivot Mentor-00009672. B.Y., K.J.M, and J.V.K are grateful for the tremendous support they received from the Gemini Observatory staff (especially Brian Lemaux and Joanna Thomas-Osip) to ensure the GMOS observations were quickly incorporated into the night queue and the instrumental configuration was relevant for the desired science goals. We thank IRTF director J. Rayner for allocating DD time, and Mike Kelii and Dwight Chan for repairing SpeX at the summit on Sunday evening in time for our observations. Finally, we thank Zahed Wahhaj for helpful input and stimulating conversations.

\end{acknowledgments}

\begin{contribution}

B.Y.\ Conducted the IRTF observations, reduced the Gemini and IRTF data, wrote large portions of the manuscript.

K.M.\ PI on the Gemini observations, secured the DD time on the IRTF. Wrote part of the manuscript and contributed to a critical review of the manuscript

M.C.\ Instrumental to successfully executing the IRTF observations. Reviewed the manuscript.

Z.R.N.\ Performed the spectral modeling and contributed to the manuscript review.

J.V.K\ Contributed to the Gemini observations planning and configuration, wrote part of the manuscript, and contributed to the manuscript review.

\end{contribution}

\bibliography{3I-Gem-IRTF,ref-OLD}{}
\bibliographystyle{aasjournal}

\end{CJK*}
\end{document}